\documentclass[preprint,showpacs,showkeys,floatfix,preprintnumbers,amsmath,amssymb]{revtex4}

\usepackage{graphics}
\usepackage{graphicx}
\usepackage{amssymb}
\usepackage{hyperref}
\usepackage{color}

\begin{document}

\bibliographystyle{apsrev}

\title{Effect of oxygen-doping on Bi$_{2}$Sr$_{2}$Ca$_{2}$Cu$_{3}$O$_{10 + \delta}$
vortex matter: Crossover from electromagnetic to Josephson
interlayer coupling}

\author{A. Piriou}%
 \email{Alexandre.Piriou@physics.unige.ch}
\author{Y. Fasano}
\author{E. Giannini}
\author{\O. Fischer}

\affiliation{%
D\'epartement de Physique de la Mati\`ere Condens\' ee,
Universit\'e de Gen\`eve, 24 Quai Ernest-Ansermet, 1211 Geneva,
Switzerland
}%

\date{\today}

\begin{abstract}

We study the effect of oxygen-doping on the critical temperature,
$T_{\rm c}$, the vortex matter phase diagram and the nature of the
coupling mechanism between the Cu-O layers in the three-layer
Bi$_{2}$Sr$_{2}$Ca$_{2}$Cu$_{3}$O$_{10 + \delta}$ (Bi-2223)
compound. Contrary to previous reports, in the overdoped (OD)
regime we do find a variation of $T_{\rm c}$ with increasing the
oxygen partial-pressure of the post-annealing treatment. This
variation is less significant than in the bi-layer parental
compound Bi$_{2}$Sr$_{2}$CaCu$_{2}$O$_{10 + \delta}$ (Bi-2212) and
does not follow the universal $T_{\rm c}$ \textit{vs.} $\delta$
relation. Magnetic  measurements reveal that increasing $\delta$
enlarges the field and temperature stability of the Bragg glass
phase. These findings imply that the interlayer coupling between
Cu-O layers enhances with $\delta$. The anisotropy parameter
estimated from directional first-penetration field measurements
monotonously decreases from 50 in the underdoped (UD) to 15 in the
OD regimes. However, the in-plane penetration depth presents a
boomerang-like behaviour with $\delta$, reaching its minimum value
close to optimal doping. These two facts lead to a crossover from
a Josephson(OD) to electromagnetic(UD)-dominated coupling of
adjacent Cu-O layers in the vicinity of optimal doping.

\end{abstract}



\maketitle

\section*{Introduction}

Understanding the extremely anisotropic  electronic and magnetic
properties of layered cuprates  requires to unveil the nature of
the coupling mechanism between the superconducting Cu-O layers. In
these materials these layers are composed of a stack of Cu-O
planes. The interlayer coupling between adjacent Cu-O layers is
produced by Josephson tunnelling and by electromagnetic
interactions between the supercurrents lying in the  layers
\cite{Blatter94a,Koshelev98a}. The family of Bi-based compounds
presents a weak interlayer coupling between the Cu-O layers
\cite{Kishio96a} resulting in both terms being comparable. The
possibility of the electromagnetic interaction being dominant over
the Josephson one was thoroughly studied
\cite{Kishio96a,Villard98a,Correa01a} for the bi-layer
Bi$_{2}$Sr$_{2}$CaCu$_{2}$O$_{8 + \delta}$ (Bi-2212) compound
which presents one of the weakest interlayer coupling among
cuprates \cite{Kishio96a}. In this work we address this issue in
the less studied three-layer
Bi$_{2}$Sr$_{2}$Ca$_{2}$Cu$_{3}$O$_{10 + \delta}$ (Bi-2223)
material by tuning the  oxygen concentration  in a single sample.
Such studies in this compound are lacking due to the difficulty in
obtaining pure and high-quality macroscopic crystals
\cite{Giannini04a} for reliable magnetic measurements. This
compound presents the highest critical temperature at optimal
doping (OPT) among the Bi-based cuprates, $T^{\rm OPT}_{\rm c} =
110.5$\,K.

The crystal structure of Bi-2223 is composed of insulating
(blocking) layers intercalating in the $c$-axis direction with
superconducting Cu-O layers. The spacing between the layers,
formed by three Cu-O planes, is $s=18$\,\AA\ for optimally-doped
samples \cite{Giannini08a}. The interlayer coupling is inversely
proportional to the electronic anisotropy, $\gamma = \sqrt{m_{\rm
c}/m_{\rm ab}}$, given by the ratio between the effective masses
in the $c$-axis and in the $ab$-plane \cite{Blatter94a}. As
mentioned, this coupling is produced by two different mechanisms.
The first, the Josephson coupling, consists of the hole-tunnelling
from one Cu-O layer to the adjacent, through the blocking layer.
This mechanism is dominant when $s \gamma < \lambda_{\rm ab}(0)$
\cite{Koshelev98a}, where $\lambda_{\rm ab}(0)$ is the in-plane
penetration depth. The second
 mechanism arises from the electromagnetic interaction
between the supercurrents located in adjacent Cu-O layers
 and is, on the contrary, dominant when
 $s \gamma > \lambda_{\rm ab}(0)$ \cite{Koshelev98a}. The relevance of these two
mechanisms depends on the hole-carrier concentration and on the
degree of charge-transfer between the Cu-O layers. One way to
modify these magnitudes is to over-dope (OD) or under-dope (UD)
with respect to the optimal oxygen content ($\delta =
\delta_{OPT}$). Extra oxygen atoms will occupy vacancies in the
blocking layers.

Oxygen-doping produces  quantitative changes in the vortex phase
diagram. Flux lines in the  extremely anisotropic  Bi-based
cuprates are composed of a stack of poorly-coupled pancake
vortices located in the Cu-O layers \cite{Blatter94a}. This weak
coupling leads to the liquid vortex phase spanning a considerable
fraction of the $H-T$ phase diagram \cite{Pastoriza94a}. On
cooling at low magnetic fields vortex matter undergoes a
first-order solidification transition   at $T_{\rm m}$
\cite{Pastoriza94a,Zeldov95b}. On further cooling the vortex
magnetic response becomes irreversible since pinning sets-in at a
temperature $T_{\rm IL}(H) \lesssim T_{\rm m}(H)$, the so-called
irreversibility line. The phase stable at low temperatures, the
Bragg glass \cite{Nattermann90a,Giamarchi94a}, exhibits
quasi-crystalline order. With increasing magnetic field the vortex
structure transforms into a topologically disordered  glass
\cite{Fisher91a} through a first-order phase transition known as
the second-peak line, $H_{\rm SP}$
\cite{Cubbit93a&Zeldov95a,Khaykovich96a}. The vortex glass melts
with increasing temperature through a second-order phase
transition  \cite{Khaykovich97a}. In the case of Bi-2212, the
Bragg glass phase  spans up to higher temperatures and fields with
increasing oxygen concentration
\cite{Kishio94a&others,Khaykovich96a,Correa01a}, which is
consistent with a decrease of $\gamma$  with $\delta$. Introducing
extra oxygen atoms also affects the pinning potential landscape
\cite{Ooi98a} Nevertheless, changes in the vortex phase diagram
provide information on the evolution of the interlayer coupling
with doping. In this work we report on the doping-dependence of
the critical current density, $J_{\rm c}(T,H)$, and on the
evolution of  $H_{\rm SP}(T)$, $H_{\rm IL}(T)$, and $T_{\rm
0D}(H)$, the zero-dimensional pinning temperature at which pancake
vortices of every Cu-O layer are individually pinned. We find that
changes in the oxygen-doping level of Bi-2223 affect the vortex
phase diagram in a similar manner than in Bi-2212. However, since
Bi-2223 is three times less anisotropic than Bi-2212, significant
differences are also detected.

In the case of the bi-layer compound, increasing the oxygen
concentration results in a monotonous reduction  of $\gamma$
\cite{Kishio96a,Correa01a}, and of the $c$-axis lattice parameter
\cite{Correa01a}, and consequently $s$. However, $\lambda_{\rm ab}
(0)$ evolves in a non-monotonous way with $\delta$, being minimum
in the slightly OD regime
\cite{Villard98a,Correa01a,Uemura89&91,Niedermayer93a}. Therefore,
given that in Bi-based compounds $s \gamma \sim \lambda_{\rm ab}$,
a crossover from electromagnetic to Josephson-dominated coupling
is quite likely when decreasing $\delta$. Evidence of this
crossover was indeed reported in the case of Bi-2212
\cite{Correa01a}. Nevertheless, as the anisotropy of Bi-2223 is
smaller than that of Bi-2212, in the three-layer compound the
Josephson coupling might be dominant in a wider $\delta$ region.
In this work we address the question whether a crossover from
Josephson to electromagnetic coupling is plausible in Bi-2223. We
find that when decreasing $\delta$, on top of increasing the
magnitude of the electronic anisotropy, this crossover takes place
in the vicinity of optimal doping.

\section*{Crystal-growth and oxygen-doping}

\begin{figure}
\begin{center}
\includegraphics[angle=-90,width=\textwidth]{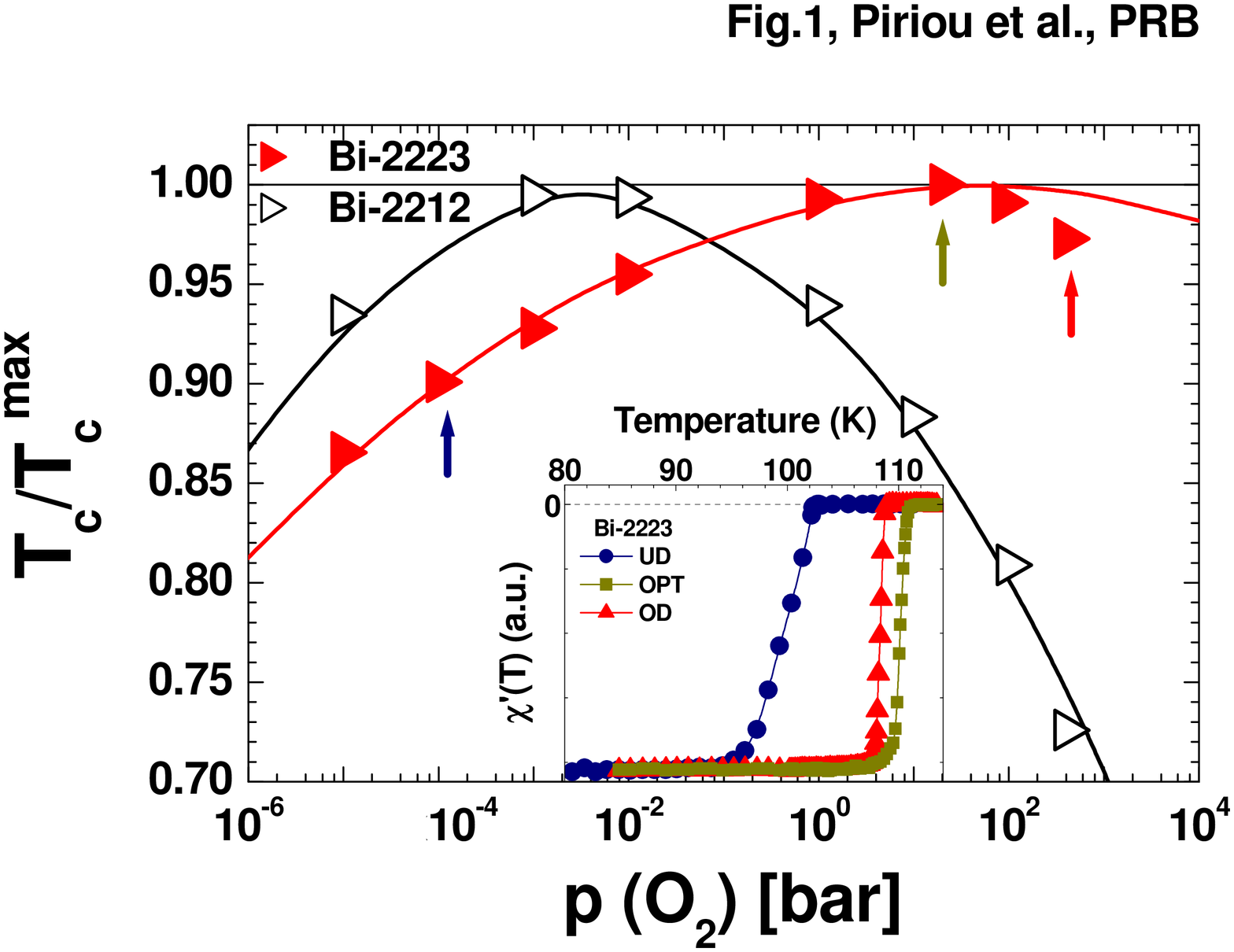}
\caption{Normalized critical temperature as a function of
post-annealing oxygen pressure $p(O_{2})$ for Bi-2223 and Bi-2212.
The maximum values of critical temperatures are $(90.5 \pm 0.5)$
and $(110.5 \pm 0.5)$\,K for Bi-2212 and Bi-2223, respectively.
 The lines are fits of the data with the universal relation
 $T_{\rm c}= T_{\rm c}^{max} (1 - 82.6(\delta
- 0.27)^2)$ \cite{Allgeier90a,Presland91a,Zhao98a} with $\delta =
0.011 \ln{p(O_{2})} + 0.3$ \cite{Zhao98a} for Bi-2212 and $\delta
= 0.006 \ln{p(O_{2})} + 0.26$ for Bi-2223. Every point in
$p(O_{2})$ corresponds to at least ten samples and the  size of
the points includes the dispersion in critical temperature. The
arrows indicate the different doping levels tuned in the sample
used to study the doping-dependence of the Bi-2223 vortex phase
diagram. Insert: Real part of the susceptibility as a function of
temperature for the same sample of Bi-2223 in the UD ($T_{\rm
c}(100 \pm 2)$\,K), OPT ($T_{\rm c}(110.5 \pm 0.5)$\,K) and  OD
($T_{\rm c}(107 \pm 0.5)$\,K) regimes. The $\chi^{'}(T)$
measurements were performed with an AC field of 0.1\,Oe in
magnitude and 970\,Hz in frequency. \label{fig:figure1}}
\end{center}
\end{figure}

Bi-2223 and Bi-2212 crystals were grown by means of the
travelling-solvent floating-zone method  as described in
Ref.\,\cite{Giannini04a}.  The structural and superconducting
properties of the crystals used in this work are reported in
Refs.\,\cite{Giannini08a,Giannini04b}. X-Ray diffraction
measurements have revealed the phase purity and the high
crystalline order of our samples. In the case of the Bi-2223
samples its quality has allowed the refinement of the crystal
structure from single-crystal X-ray diffraction
data\,\cite{Giannini08a}. Electrical resistivity and magnetic
susceptibility measurements reveal a single and sharp
superconducting transition (examples are shown in the insert of
Fig.\,\ref{fig:figure1}). Considering the measurements resolution,
these results indicates that, if present, Bi-2212 intergrowths
represent less than 1 \% of the volume of the samples.

In order to tune and to homogenize the hole concentration the
crystals were annealed during 10 days at 500$^{\circ}$C, under
various oxygen partial pressures, $p(O_{2})$. In
Fig.\,\ref{fig:figure1} we compare the dependence of $T_{\rm c}$
on $p(O_{2})$ for both Bi-2212 and Bi-2223 crystals. Every point
results from an average over at least 10 samples annealed in the
same conditions. At every $p(O_{2})$ value, the critical
temperature and the transition width are the same within the
 error, as measured by means of AC susceptibility and
DC magnetization.  The reliability of these results allows us to
consider the critical temperature as a measure of the macroscopic
oxygen content. The transition temperature $T_{\rm c}$ is taken as
the temperature at which $\partial \chi^{'} /\partial T$  and
$\partial M /\partial T$ exhibit a peak. The transition widths are
estimated as the FWHM of these peaks and range from 0.6 to 2\,K.
This is an indication that in our crystals the oxygen distribution
is rather homogeneous.

In the case of Bi-2212 samples the $T_{\rm c}$ \textit{vs.}
$p(O_{2})$ data are in quantitative agreement with the existing
literature \cite{Kishio96a} and do not deserve any further
comment. However, the results for Bi-2223 are in  qualitative
disagreement with previous reports \cite{Fujii02a,Liang04a}. As in
the case of these works, our OPT Bi-2223 samples, with $T_{\rm c}
= (110.5 \pm 0.5)$\,K,  are obtained at quite a high partial
pressure, $p(O_{2}^{OPT}) = 20$\,bar. For higher annealing
pressures the samples become overdoped. Fig.\,\ref{fig:figure1}
shows that $T_{\rm c}$ does  change in the OD regime, decreasing
from $(110.5 \pm 0.5$ to $(107.0 + 0.5)$\,K. This is in contrast
with the works of Fujii \textit{et al.} \cite{Fujii02a} and Liang
\textit{et al.} \cite{Liang04a} which report a  $T_{\rm c}$
\textit{vs.} $p(O_{2})$ plateau in the OD regime. The authors
\cite{Fujii02a,Liang04a} claim that this plateau results from a
different doping level of the inner and outer planes of the Cu-O
layers, the inner being less doped \cite{Trokiner91a}. This
interpretation is based on a theoretical model that assumes a
strong interlayer coupling \cite{Kivelson02a}.

In spite of the inner and outer Cu-O planes having a different
doping level \cite{Trokiner91a}, other factors might be at the
origin of the $T_{\rm c}$-plateau reported in
Refs.\,\cite{Fujii02a,Liang04a}. As a matter of fact, an
inhomogeneous oxygen distribution throughout the sample can be
responsible for an apparent insensitivity of $T_{\rm c}$ to oxygen
doping. An inhomogeneous oxygen distribution is also compatible
with the detected decrease of the \textit{c}-axis lattice
parameter \cite{Fujii02a} and the enhancement of the vortex
irreversibility field \cite{Liang04a} with increasing $\delta$.
Crystals exhibiting a broader superconducting transition than
ours, like the ones reported in Refs.\,\cite{Fujii02a,Liang04a},
might present an inhomogeneous oxygen spatial-distribution. In
such crystals the onset of the superconducting transition would be
determined by the
 OPT domains, whereas the shift of the x-ray
reflections and the enhancement of $H_{\rm IL}(T)$ would be
dominated by the OD ones. The  narrower superconducting
transitions measured in our crystals indicate that their spatial
oxygen distribution is more homogeneous and as a consequence the
detected decrease of $T_{\rm c}$ in the OD regime  is not hidden
by inhomogeneities of the samples.

The most striking result of Fig.\,\ref{fig:figure1} is that a
larger oxygen partial-pressure is required in order to reach
optimal doping in Bi-2223 than in Bi-2212. Our results for Bi-2212
samples are in agreement with the established bell-shaped
dependence of $T_{\rm c}$ with oxygen concentration
\cite{Allgeier90a,Presland91a,Zhao98a}, see fits of
Fig.\,\ref{fig:figure1}. On the contrary, in the case of Bi-2223
the curve resulting from adjusting the UD  data with the universal
$T_{\rm c}$ \textit{vs.} doping ($p(O_{2})$) relation
\cite{Allgeier90a,Presland91a}  does not fit the points in the OD
regime. Seminal works on three-layer cuprates already reported
that $T_{\rm c}$ follows the universal law  in the UD region
\cite{Presland91a,Schilling89a}. Our results in Bi-2223 confirm
this  and also indicate that in the OD regime of three-layer
compounds the universal law \cite{Allgeier90a,Presland91a} is not
fulfilled as in the case of single- and bi-layer cuprates. This
phenomenology can be related to the different doping level of the
outer and inner Cu-O planes.

\section*{Doping-dependence of the vortex phase diagram}

As mentioned, to study the effect of hole-doping on the vortex
phase diagram of Bi-2223 the oxygen content was tuned in the same
sample. The results presented in this section correspond to doping
levels in the UD ($T_{\rm c} = (100 \pm 2)$\,K), OPT ($T_{\rm c} =
(110.5 \pm 0.5)$\,K) and OD ($T_{\rm c} = (107.0 \pm 0.5)$\,K)
regimes.

\begin{figure}[ttt]
\includegraphics[width=0.8\textwidth]{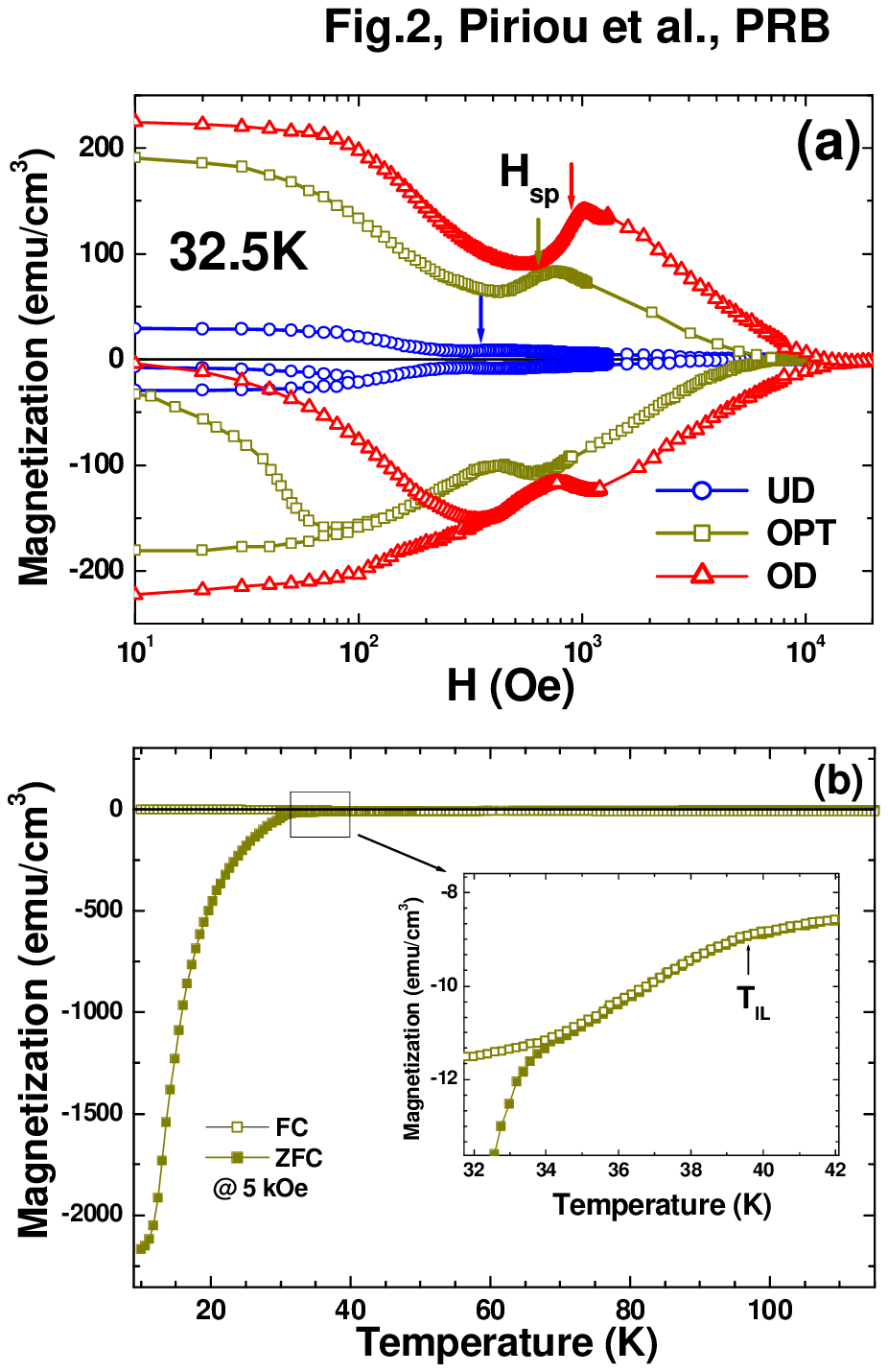}
\caption{ (a) Magnetization loops at 32.5\,K for the same sample
for UD ($T_{\rm c} = (100 \pm 2)$\,K), OPT ($T_{\rm c} = (110.5
\pm 0.5)$\,K) and OD ($T_{\rm c} = (107.0 \pm 0.5)$\,K) oxygen
concentrations. The  second peak field $H_{\rm SP}(T)$ is
indicated with arrows. (b) Field-cooling (FC) and
Zero-field-cooling (ZFC) magnetization measurements for the OPT
regime at an applied field of 5\,kOe. The onset-temperature of the
irreversible magnetic behavior at 5\,kOe, $T_{\rm IL}$, is
indicated. \label{fig:figure2}}
\end{figure}

The vortex phase diagram was studied  by means of bulk magnetic
measurements using a Quantum-design SQUID magnetometer and a PPMS
measurement system. In particular, we report on the evolution of
the irreversibility line, $H_{\rm IL}$(T), the order-disorder
$H_{\rm SP}$(T) transition line,  associated with a peak in the
critical current \cite{Cubbit93a&Zeldov95a,Khaykovich96a}; and the
zero-dimensional pinning line $T_{\rm 0D}$
\cite{Niderost96a,Goffman98a}. These lines were obtained from
magnetization \emph{vs.} temperature measurements, $M(T)$,
following zero-field- (ZFC) and field-cooling (FC) paths, and from
hysteresis magnetization loops, $M(H)$. For these measurements the
field was applied out-of-plane, i.e. $H \parallel c$-axis. The
typical field and temperature sweep-rates in $M(H)$ and $M(T)$
measurements were of 10$^{-2}$\,Gauss/sec  and 10$^{-3}$\,K/sec,
respectively. Figure\,\ref{fig:figure2} presents typical
magnetization data for doping levels corresponding to the UD, OPT
and OD regimes.  The superconducting transition in each case is
shown in the insert of Fig.\,\ref{fig:figure1}.

\begin{figure}[ttt]
\includegraphics[angle=-90,width=\textwidth]{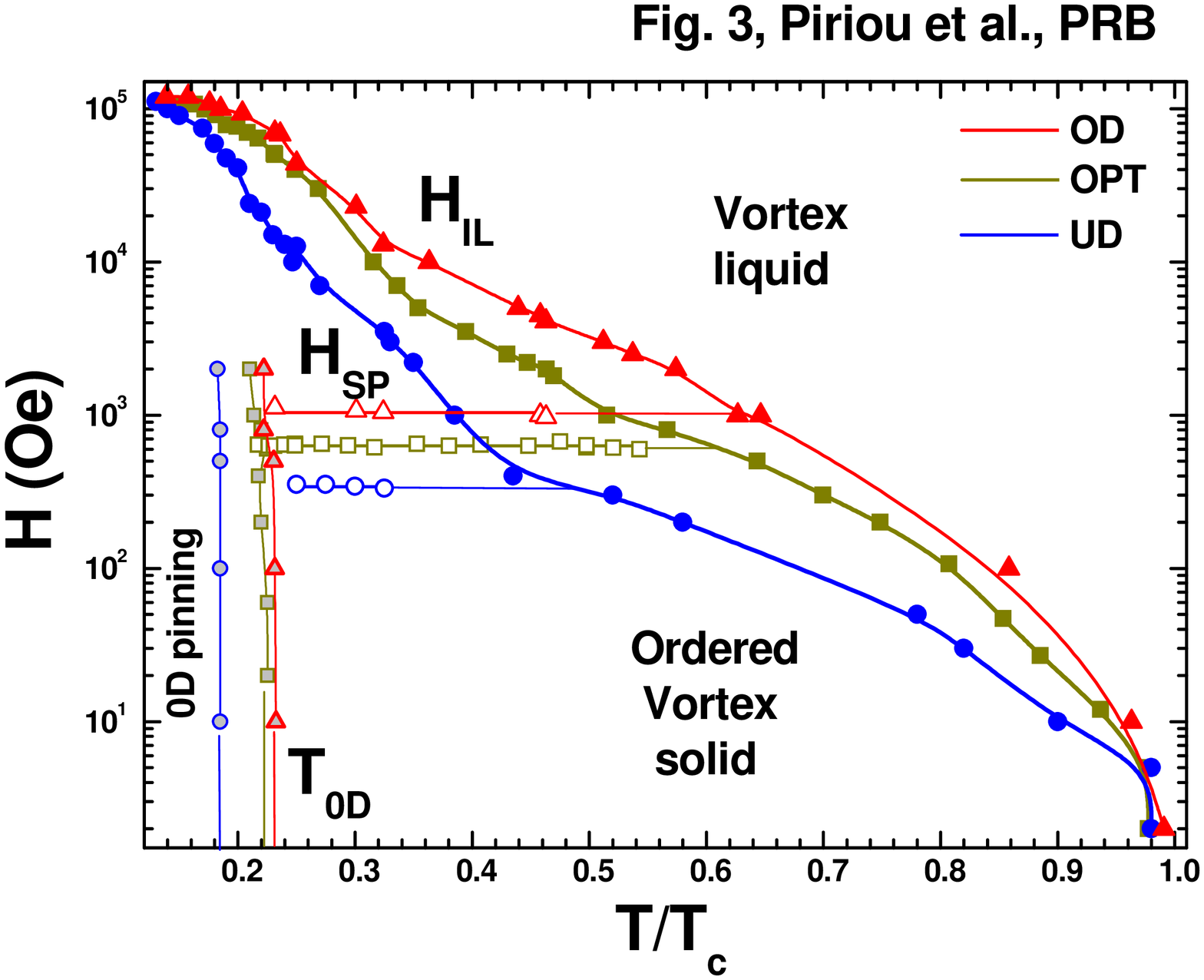}
\caption{ Vortex phase diagram for the same sample in UD
($\bigcirc$ ), OPT ($\square$ ) and extremely OD ($\triangle$ )
regimes. The irreversibility, $H_{\rm IL}(T)$ (full symbols),
second peak $H_{\rm SP}(T)$ (open symbols) and zero-dimensional
pinning $T_{\rm 0D}(H)$ (gray-filled symbols) lines are shown. The
error in magnetic field and temperature of the measured points is
included in their size. \label{fig:figure3}}
\end{figure}

The onset of the irreversible magnetic response was obtained from
field- and zero-field-cooling $M(T)$ curves as the kink detected
at a temperature $T_{\rm IL}(H)$ higher than that where both
branches merge \cite{Schilling92a}, see Fig.\,\ref{fig:figure2}
(b).   The hole-doping dependence of the irreversibility line  is
shown in the $H \, vs. \,T/T_{\rm c}$ phase diagram of
Fig.\,\ref{fig:figure3}. At any reduced temperature, the
irreversibility field is enhanced with increasing the
oxygen-doping level.   The evolution of $H_{\rm IL}(T)$ with
doping for the bi-layer compound \cite{Correa01a} and data
previously reported for Bi-2223 \cite{Piriu07a,Liang04a} follow
the same trend.

The enlargement of the solid vortex phase is consistent with an
enhancement of the interlayer coupling by increasing oxygen
concentration, and consequently a reduction of the anisotropy
parameter. Theoretical studies indicate that the melting
\cite{Blatter96a} and decoupling \cite{Daemen93a} lines shift
towards higher temperatures as soon as a small Josephson coupling
between the layers is considered on top of electromagnetic
interactions. In the case of platelet-like samples, such as the
ones studied in this work, for Bi-2212 at low fields the
irreversibility coincides with the melting and decoupling lines
\cite{Pastoriza94a,Goffman98a}. Therefore, the enhanced stability
of the solid vortex phase shown in Fig.\,\ref{fig:figure3} can be
associated with an increasingly relevant role of the Josephson
coupling when raising the doping level. This is quite likely since
 $s \gamma \gtrsim \lambda_{\rm ab}$ for the OPT regime of
Bi-2223. This issue will be further discussed in the next section.

The transition line $H_{\rm SP}(T)$ is manifested as a peak-valley
structure in  $M(H)$ curves, as evident in the  magnetization
loops of Fig.\,\ref{fig:figure2} (a). We considered $H_{\rm
SP}(T)$ to be the field at which $M$ has an inflection point
between the peak and the valley (indicated by arrows in the
figure), averaging the values for the two branches of the loop.
The phase diagram of Fig.\,\ref{fig:figure3} shows that the
second-peak line is roughly temperature-independent, as also found
in Bi-2212 \cite{Khaykovich96a,Goffman98a,Correa01a}. Theoretical
works considering the $H_{\rm SP}(T)$ line as an order-disorder
phase transition between the low-field Bragg glass and the
high-field vortex glass
\cite{Giamarchi94a,Ertas96a&Giamarchi97a&Vinokur98a&Kierfeld98a}
estimate $H_{\rm SP}(T)$  as the field where the elastic energy of
the vortex structure equals the  pinning energy. These two energy
terms depend on the penetration depth, coherence length and
anisotropy of the material, as well as on the pinning parameter
\cite{Giamarchi94a}. Therefore,  the temperature evolution of
$H_{\rm SP}$ is determined by the temperature dependence of
$\lambda_{\rm ab}$ and $\xi_{\rm ab}$ \cite{Giller99a}.
Figure\,\ref{fig:figure3} shows that for the three doping levels
studied $H_{\rm SP}$ is detected only in the temperature range
$0.2 \lesssim T/T_{\rm c} \lesssim 0.55$. Considering the
two-fluid expression for $\lambda_{\rm ab}$ and $\xi_{\rm ab}$
\cite{Blatter94a}, in this temperature range both magnitudes vary
within $3$\,\%. This minimal variation and the important
anisotropy \cite{Giller99a} of the material are responsible for
the observed roughly temperature-independent $H_{\rm SP}$.

The temperature-averaged $H_{\rm SP}$ increases with doping: $(340
\pm 20)$\,Oe for the UD, $(630 \pm 40)$\,Oe for the OPT and $(1040
\pm 60)$\,Oe for the OD regimes. The same qualitative behavior was
reported in Bi-2212
\cite{Kishio94a&others,Khaykovich96a,Correa01a} and other cuprates
\cite{Hardy94a&Shibata02a&Masui04a}. The evolution of $H_{\rm
SP}(T)$ with doping is consistent with an enhancement of coupling
between the Cu-O planes with increasing oxygen concentration, as
also suggested by the doping dependence of $H_{\rm IL}(T)$
\cite{Koshelev98a}. This point will be further discussed in the
next section.

The zero-dimensional pinning line, $T_{\rm 0D}$, that separates
the regime where individual pancake vortices are pinned ($T <
T_{\rm 0D}$) from that where vortex lines are pinned individually,
can be obtained from critical current density \textit{vs.}
temperature curves $J_{\rm c} (T)$ at a given magnetic field
\cite{Correa01b}. In the low-temperature zero-dimensional pinning
regime not only the vortex-pinning interaction overcomes the
vortex-vortex interaction but also pancake vortices belonging to
the same vortex but to different Cu-O layers are pinned
individually \cite{Blatter94a}. Therefore at temperatures $T <
T_{\rm 0D}$ the critical current should be field independent at
low fields. This is a consequence of the $c$-axis Larkin
correlation length, $L^{\rm c}_{\rm c} \sim (r_{p}/\gamma) (J_{\rm
0}/J_{\rm c})^{1/2}$ \cite{Larkin79a}, becoming smaller than the
Cu-O layers spacing $s$ \cite{Blatter94a}. In the last expression
$r_{p} \sim \xi$ is the typical pinning range and $J_{\rm 0}= 4c
\Phi_{\rm 0}/ 12 \sqrt{3} \pi \lambda^{2}_{\rm ab} \xi_{\rm ab}$
is the depairing current density. Therefore $T_{\rm 0D}(T)$ can be
estimated as the temperature at which $J_{\rm c}(T)$ presents a
kink and a steeper increase in the low-temperature region
\cite{Correa01b}. In this work we consider the same criterion to
estimate $T_{\rm 0D}$.

\begin{figure}[ttt]
\includegraphics[width=\textwidth]{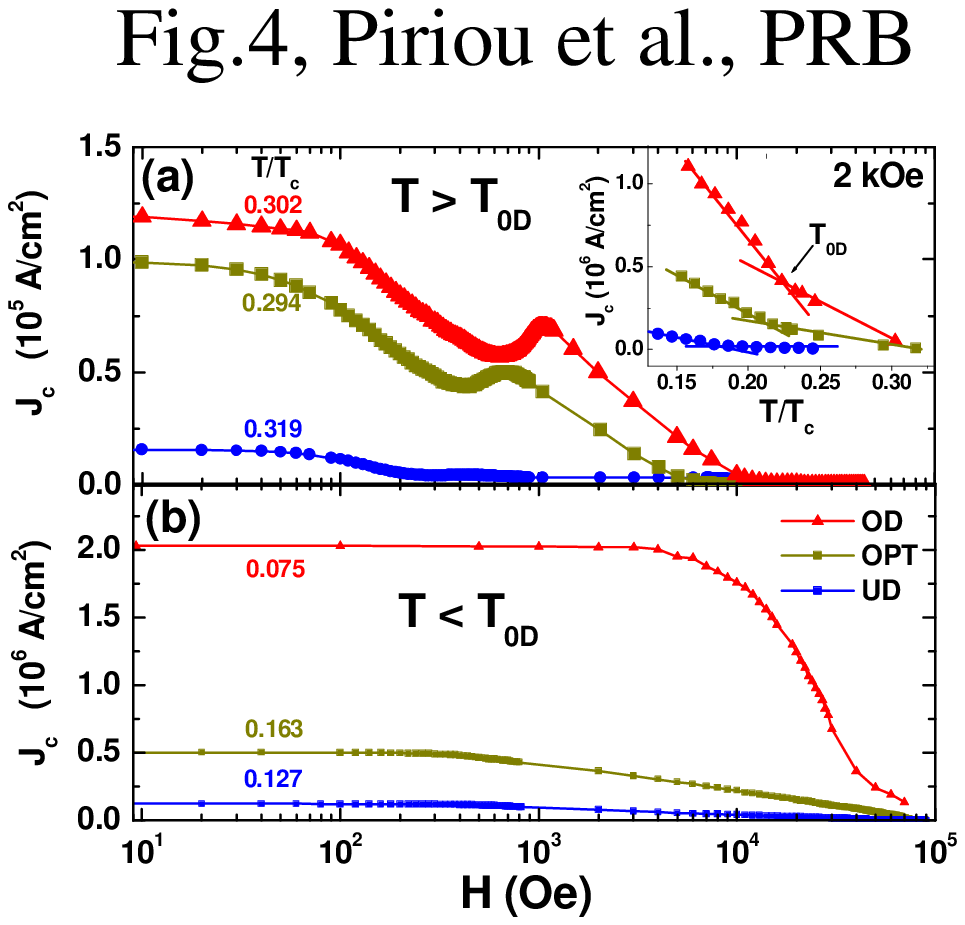}
\caption{ Critical current density as a function of magnetic field
, $J_{\rm c}(H)$, for (a) $T > T_{\rm 0D}$ and (b) $T < T_{\rm
0D}$. Insert: Curves of $J_{\rm c}$ \textit{vs.} reduced
temperature obtained from the width of the magnetization loop at
40\,Oe (see text). The arrows indicate the kink in $J_{\rm c}$
considered as the temperature $T_{\rm 0D}$ at which the
zero-dimensional pinning sets in. \label{fig:figure4}}
\end{figure}

The critical current is obtained from magnetization loops measured
at different temperatures as the examples shown in
Fig.\,\ref{fig:figure2} (a). Considering the Bean model
\cite{Bean64a}, at a given temperature $J_{\rm c} (T,H) \sim
(3c/2R) \Delta M (T,H)$, where $\Delta M (T,H)$ is the separation
between the two branches of the magnetization loop at a field $H$,
$c$ is the speed of light and $R$ is the radius of an equivalent
cylindrical sample \cite{Bean64a}. Figure\,\ref{fig:figure4}
reveals that the critical current of Bi-2223 is one order of
magnitude larger than that of Bi-2212 \cite{Correa01b} at similar
reduced temperatures and fields. From this figure it is also
evident that at low temperatures the critical current is
field-independent at low fields, which is a fingerprint of the
individual pinning  regime (either of flux lines or of individual
pancakes). The insert of Fig.\,\ref{fig:figure4} shows typical
$J_{\rm c}(T/T_{\rm c})$ curves at an applied field of 40\,Oe for
the UD, OPT and OD regimes. The kink in critical current
identified as $T_{\rm 0D}$ is indicated with arrows. Considering
that for the OPT regime of Bi-2223 $\xi \sim 10$\,\AA \cite{Nate},
$J_{\rm 0} = 6.15 \cdot 10^{8}$\,A/cm$^2$, $\gamma = 27 \pm 4$
(see data shown in the next section), and $J_{\rm c} (T_{\rm 0D})
= 1.81 \cdot 10^5$\,A/cm$^2$, a value of $L^{\rm c}_{\rm c} =
24-18$, close to $s$ is estimated at $T=T_{\rm 0D}$. Similar
values for $L^{\rm c}_{\rm c}$ are obtained for the UD and OD
regimes.

For every doping level the zero-dimensional-pinning line is
roughly field-independent, as also found in Bi-2212
\cite{Goffman98a,Correa01a}. Since $T_{\rm 0D}$ is the temperature
at which the individual pinning of pancake vortices sets in, then
$J_{\rm c}(T)$ is almost field-independent for $T \leq T_{\rm
0D}$. Therefore, $L^{\rm c}_{\rm c}\sim (r_{p}/\gamma) (J_{\rm
0}/J_{\rm c}(T_{\rm 0D}))^{1/2}$ equals the Cu-O layers spacing at
a temperature $T_{\rm 0D}$ that is roughly field-independent. The
location of $T_{\rm 0D}$ is however doping dependent. The
zero-dimensional pinning line is placed at $(18.5 \pm 0.5)$\,K for
the UD, $(24 \pm 1)$\,K for the OPT and $(25 \pm 1)$\,K for the OD
regimes. A study
 on the OPT and OD regime of Bi-2212 samples reports
that $T_{\rm 0D}$ does not significantly change with doping
\cite{Correa01b}. However, as the magnitude of $J_{\rm c}(T_{\rm
0D})$ is in direct relation with $T_{\rm 0D}$ (see
Fig.\,\ref{fig:figure4}), if $\gamma$ decreases then, in order to
fulfill $L^{\rm c}_{\rm c}=s$, $J_{\rm c}(T_{\rm 0D})$ and
therefore $T_{\rm 0D}$ have to increase.  The difference between
our results in Bi-2223 and those reported for Bi-2212
\cite{Correa01b} might originate in the fact that in the latter
study the anisotropy changes are of the order of 30\% whereas in
our case they are 5 times greater than this value (see the next
section for detailed data in this respect). Therefore, an
increment of the zero-dimensional temperature with doping is in
agreement with the enhancement of interlayer coupling suggested by
the doping evolution of $H_{\rm IL}(T)$ and $H_{\rm SP}(T)$.

One interesting result of the vortex phase diagrams shown in
Fig.\,\ref{fig:figure3} is that for $T<T_{\rm 0D}$ the
order-disorder second peak line is no longer detected within our
experimental resolution. Similar findings were observed in Bi-2212
samples \cite{Goffman98a,Correa01b}. The vortex glass phase
located at $H > H_{\rm SP}$ presents no vortex phase correlation
in the $c$ direction \cite{Goffman98a}, whereas the
zero-dimensional pinning region is correlated in the $ab$ plane as
well as in the perpendicular direction \cite{Correa01b}.
Therefore, the suppression of the vortex phase correlation along
the $c$ direction takes place only when $L^{\rm c}_{\rm c}$
exceeds the spacing between the Cu-O layers. This suggests that
the establishment of a zero-dimensional pinning regime inhibits
the vortex structure instability associated with the
order-disorder phase transition detected at $T > T_{\rm 0D}$.

In summary, the doping dependence of the transition lines $H_{\rm
IL}(T)$ and $H_{\rm SP}(T)$ and the crossover temperature $T_{\rm
0D}$ are in agreement with an enhancement of the coupling between
Cu-O layers with increasing doping. The decrease of $\gamma$ on
overdoping suggested by the results presented in this section
implies that the Josephson coupling term between adjacent layers
becomes increasingly relevant. Therefore, the scenario of a
crossover from electromagnetic(UD) to Josephson(OD)-dominated
interlayer coupling  is quite likely. However, a possible
non-monotonous evolution of $\lambda_{\rm ab}$ with doping has to
be explored in order to differenciate between this scenario and
the situation where the interlayer coupling increases but
continues to be electromagnetic in nature when overdoping.

\section*{Crossover from electromagnetic to Josephson coupling when overdoping}

In order to study the  nature of the coupling between the Cu-O
layers of Bi-2223 as a function of doping we quantitatively
compared the characteristic lengths $s \gamma$ and $\lambda_{\rm
ab}(0)$. In the case of Bi-2212 a crossover from electromagnetic
($s \gamma > \lambda_{\rm ab}$) to Josephson($s \gamma <
\lambda_{\rm ab}$)-dominated interlayer coupling takes place
slightly above  the OPT regime (for $T_{\rm c} \sim  0.9 T_{\rm
c}^{max}$) \cite{Correa01a}. We study this possibility in the case
of Bi-2223.

The in-plane penetration depth is extracted from
 measurements of the first penetration field
with $H$ applied perpendicular to the Cu-O planes, $H_{\rm
c1}^{\perp}$, for different temperatures in the range 35-60\,K.
The first penetration field was obtained from $M \, vs.\,H$ loops
where the magnetic relaxation at every field was measured during
one hour. This was done in order to avoid the effect of surface
and geometrical pinning barriers \cite{Niderost98a}. The effective
first penetration field for the sample was considered as that
where the magnetization shows a detectable relaxation, associated
with the entrance of the first vortex. The effect of the
demagnetizing factor of the sample estimated from the Meissner
slope was considered in order to obtain $H_{\rm c1}^{\perp}$.  The
value of $\lambda_{\rm ab} (0)$ was obtained by fitting the
temperature-dependent $H_{\rm c1}^{\perp}$ within the London
model, $H_{\rm c1}^{\perp}=(\Phi_{\rm 0}/(4 \pi \lambda_{\rm
ab}(T)^{2})\ln{(\lambda_{\rm ab}(T)/\xi_{\rm ab}(T))}$, and
considering the two-fluid model expressions for $\lambda_{\rm
ab}(T)$ and $\xi_{\rm ab}(T)$ \cite{Blatter94a}. The
zero-temperature coherence length is taken as $\xi(0) = 10$\,\AA,
as suggested by Scanning Tunnelling Microscopy measurements of the
superconducting
 density of states in the vicinity of a vortex \cite{Nate}.

\begin{figure}
\includegraphics[angle=-90,width=\textwidth]{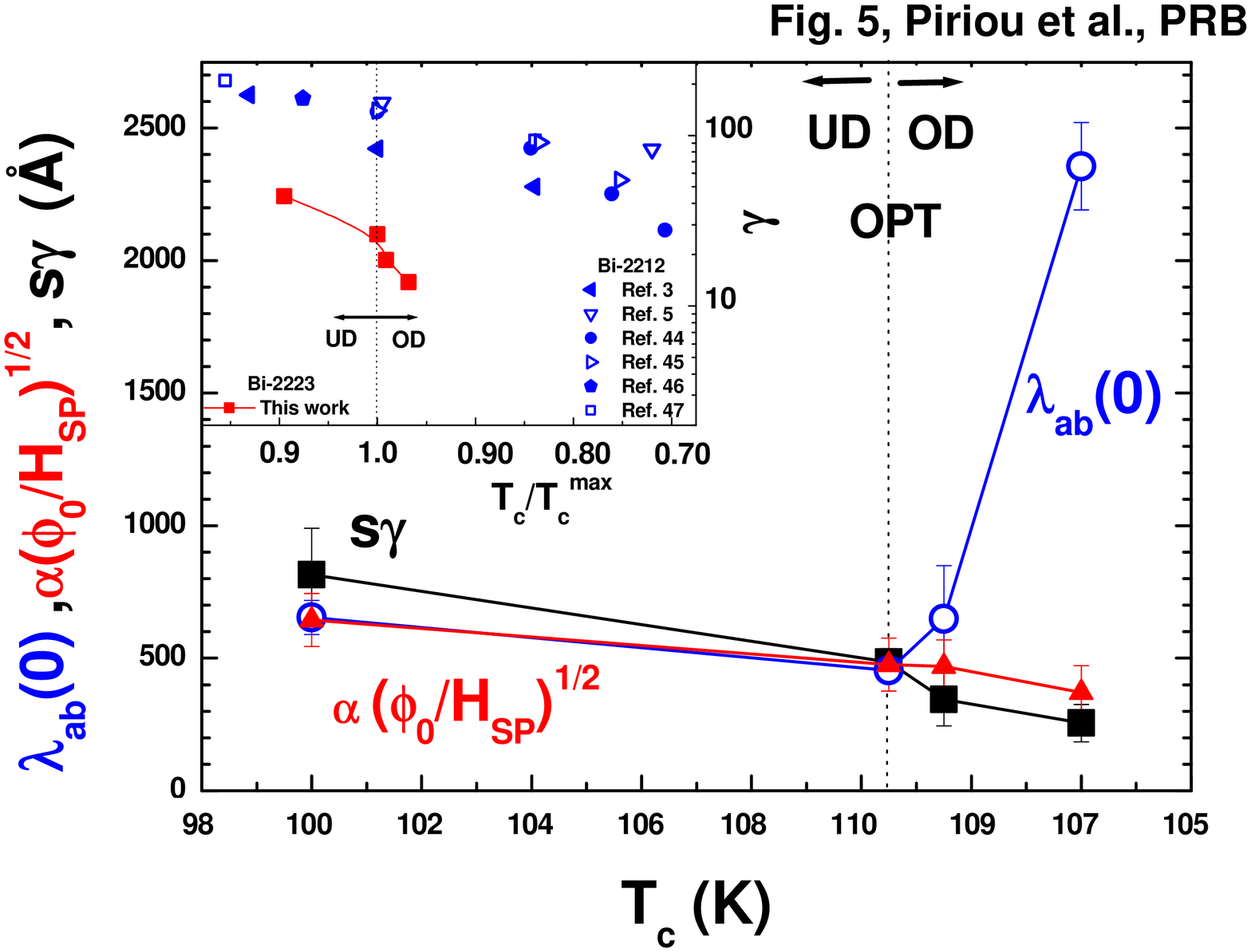}
\caption{Evolution of the in-plane penetration depth,
$\lambda_{\rm ab}(0)$, the anisotropy, $s \gamma$, and the
second-peak field, $\propto \sqrt{\Phi_{\rm 0}/H_{\rm SP}}$,  for
the UD, OPT and OD regimes of Bi-2223. Insert: Electronic
anisotropy as a function of $T_{\rm c}/T_{\rm c}^{max}$ for our
Bi-2223 sample and values reported in the literature for numerous
Bi-2212 samples
\cite{Kishio96a,Correa01a,Darminto02a,Darminto01a,Musolino04a,Nakayama00a}.
\label{fig:figure5}}
\end{figure}

The boomerang-like dependence of $\lambda_{\rm ab}(0)$ with oxygen
concentration is shown in Fig.\,\ref{fig:figure5}: it decreases
from the UD towards the OPT region and then substantially
increases in the OD regime. This non-monotonic evolution is in
agreement with results obtained in other cuprates
\cite{Niedermayer93a,Tallon95a,Villard98a}. It was originally
proposed  that  this decrease of the superfluid density in the OD
regime results from a substantial pair breaking  in spite of the
normal-state carrier concentration increasing when overdoping
\cite{Niedermayer93a} . However, Scanning Tunnelling Microscopy
data on several cuprates are at odds with this interpretation
since the superconducting quasiparticle peaks sharpen on
increasing doping \cite{Fischer07a}. Therefore, the origin of this
phenomenon remains still as an open question.

The anisotropy parameter is estimated from the first penetration
field for $H$ applied perpendicular and parallel to the Cu-O
planes since within the London approximation $\gamma =
H_{c1}^{\perp}/H_{c1}^{\parallel}$ \cite{Blatter94a}. The values
of $H_{\rm c1}^{\parallel}$ were measured by aligning the sample
with a home-made rotation system that reduces the misalignment
uncertainty to $\sim 0.5^{\circ}$. The effect of the demagnetizing
factor was corrected from the Meissner slope, giving a value in
accordance with
 the $D_{\parallel} \sim (1 - D_{\perp})/2$ relation found for platelet-like samples.
 The values of $\gamma$ presented in
 the insert of Fig.\,\ref{fig:figure5} correspond to averaging results obtained
 at different temperatures ranging between 35 and 60\,K. The error
 bar in the points represent the dispersion of values measured at
 different temperatures. The anisotropy parameter monotonically
 decreases with increasing oxygen concentration. This establish that the
 interlayer coupling is enhanced
 whith increasing the oxygen concentration, as suggested by the results presented in the previous section.
 We estimate a value
 of $\gamma = (27 \pm 3)$ for the optimally doped sample, smaller than
 previously reported \cite{Clayton04a}. It is important to point out that our estimation
 is a result of measuring the first critical field at various temperatures whereas in
 Ref.\,\cite{Clayton04a} it was estimated from measurements at a single temperature.

 The same evolution of the anisotropy parameter with oxygen-doping was
found for Bi-2212, one of the most anisotropic cuprates. The
insert of Fig.\, \ref{fig:figure5} compares the evolution of
$\gamma$ with $T_{\rm c}/T^{max}_{\rm c}$ for the bi-layer and
three-layer Bi-based compounds. Although the Bi-2212 data are
roughly three times larger than the Bi-2223 ones,  the variation
of $\gamma$ relative to $\gamma^{\rm OPT}$ is similar in both
compounds. Therefore the coupling between the Cu-O layers of
Bi-2223 is affected by oxygen-doping in a similar amount, relative
to $\gamma^{\rm OPT}$, than in Bi-2212.

Changes in the oxygen concentration affect the anisotropy
parameter mostly because the number of carriers in the charge
reservoir blocks varies, but also because the $c$-axis lattice
parameter is slightly modified. Adding extra oxygen reduces the
$c$-axis lattice parameter and therefore $s$ \cite{Correa01a}. In
the case of Bi-2212 the overdoping-induced decrease of the
$c$-axis lattice parameter was found to be of only $\sim 0.5$\,\%
\cite{Correa01a}. This change is much smaller than the error we
have in determining $\gamma$, and therefore in estimating $s
\gamma$ we consider the value $s=18$\,\AA\ measured in OPT samples
as the spacing between the Cu-O layers at all doping levels.

Another parameter that gives information on the nature of the
interlayer coupling is the doping evolution of $\sqrt{\Phi_{\rm
0}/H_{\rm SP}}$. In the case of the interlayer coupling being
dominated by the electromagnetic term,  $\sqrt{\Phi_{\rm 0}/H_{\rm
SP}} \propto \lambda_{\rm ab}$ \cite{Koshelev98a}. On the
contrary, if the Josephson interaction dominates, $\sqrt{\Phi_{\rm
0}/H_{\rm SP}} \propto s \gamma$ \cite{Koshelev98a}.
Figure\,\ref{fig:figure5} shows that $\sqrt{\Phi_{\rm 0}/H_{\rm
SP}}$ monotonously decreases with doping, and therefore
significantly deviate from the evolution of $\lambda_{\rm ab}$
with doping in the OD regime.

To summarize, Fig.\,\ref{fig:figure5} shows a comparison of the
characteristic lengthscales. In the OD regime,  $s \gamma <
\lambda_{\rm ab}(0)$ and $\sqrt{\Phi_{\rm 0}/H_{\rm SP}}$ does not
follow the trend of $\lambda_{\rm ab}(0)$ but rather that of $s
\gamma$. These findings indicate that the Josephson interaction
progressively dominates the interlayer coupling whith increasing
oxygen concentration from optimal doping. At the OPT regime $s
\gamma$  becomes of the order of the in-plane penetration depth.
Deep into the UD region $s \gamma$ slightly overcomes
$\lambda_{\rm ab}(0)$ and $\sqrt{\Phi_{\rm 0}/H_{\rm SP}}$ follows
the same trend than the penetration depth. These results
constitute evidence that in the UD regime the interlayer coupling
is dominated by the electromagnetic term. Therefore, with
increasing doping from the UD to the OD regime not only the
anisotropy decreases but the coupling between Cu-O layers changes
in nature, presenting a crossover from a electromagnetic to a
Josephson-dominated interaction. Although with our data the exact
doping level at which this crossover takes place can not be
accurately determined, it is evident that it occurs in the
vicinity of the OPT regime.

\section{Conclusions}

We provide evidence that oxygen-doping in Bi-2223 crystals
produces a non-negligible change of $T_{\rm c}$ in the OD regime,
in contrast to previous claims \cite{Fujii02a,Liang04a}. The
 changes in $T_{\rm c}/T_{\rm c}^{max}$ are non-symmetric with respect to
optimal doping: in the OD regime they are greater than in the UD
regime. Therefore, the evolution of $T_{\rm c}$ in the OD regime
of Bi-2223 does not follow the universal relation with $\delta$ as
found in many single- and bi-layer cuprates
\cite{Allgeier90a,Presland91a}.

Varying oxygen concentration affects the vortex matter phase
diagram  in a way that is consistent with an enhancement of the
Cu-O interlayer coupling with increasing $\delta$. Namely, the
Bragg glass phase spans up to higher fields ($H_{\rm SP}(T)$
increases) and temperatures ($T_{\rm IL}(H)$ and $T_{\rm 0D}(H)$
increase) when overdoping. The evolution of $H_{\rm IL}(T)$ and
$H_{\rm SP}(T)$ with doping is in agreement with results reported
for the bi-layer parental compound
\cite{Correa01a,Khaykovich96a,Kishio96a}. In the case of the
zero-dimensional pinning temperature we do detect an increase with
$\delta$ contrary to reports on Bi-2212 \cite{Correa01b}. This can
be understood considering that in our work $\delta$ is varied in a
much larger interval than in Ref.\,\onlinecite{Correa01b}.

The suggested enhancement of the interlayer coupling with
increasing $\delta$ is indisputably observed by the monotonous
decrease of anisotropy with oxygen concentration. A comparison
between the directly-measured magnitudes $s\gamma$, $\lambda_{\rm
ab}$, and $\sqrt{\Phi_{\rm 0}/H_{\rm SP}}$ reveals that a
crossover from Josephson(OD) to electromagnetic(UD)-dominated
coupling takes place around optimal doping. This crossover
originates from the non-monotonous behavior of $\lambda_{\rm ab}$
with $\delta$. Therefore, the highly-anisotropic superconducting
properties of UD Bi-2223 relies not only on a decrease of the
magnitude of the Cu-O layer interaction but also in a change of
the nature of its coupling mechanism.

The authors acknowledge G. Nieva for useful discussions and R.
Lortz for assistance in the SQUID measurements. This work was
supported by the MANEP National Center of Competence in Research
of the Swiss National Science Foundation.

%




\newpage

\newpage

\newpage

\newpage

\newpage

\end{document}